\documentclass[showpacs,aps,graphicx,twocolumn]{revtex4}
\usepackage{graphicx}

\begin{document}
\title{Distillation of logic-qubit entanglement assisted with cross-Kerr nonlinearity}

\author{Yu-Bo Sheng$^{1}$\footnote{Email address:
shengyb@njupt.edu.cn}  and Lan Zhou$^{1,2}$ }
\address{
$^1$Key Lab of Broadband Wireless Communication and Sensor Network
 Technology,
 Nanjing University of Posts and Telecommunications, Ministry of
 Education, Nanjing, 210003,
 China\\
$^2$College of Mathematics \& Physics, Nanjing University of Posts and Telecommunications, Nanjing,
210003, China\\}
\begin{abstract}
Logic-qubit entanglement has attracted much attention in both quantum communication and quantum computation. Here, we present  an efficient protocol to distill the logic-qubit entanglement with the help of cross-Kerr nonlinearity. This protocol not only can purify the logic bit-flip error and logic phase-flip error, but also can correct the physical bit-flip error completely.   We use cross-Kerr nonlinearity to construct quantum nondemolition detectors.  Our distillation protocol for logic-qubit entanglement may be useful for the practical applications in quantum information, especially in long-distance quantum communication.
\end{abstract}
 \maketitle

\section{Introduction}
Entanglement is of very importance in most quantum communication protocols. The typical quantum communication protocols such as  quantum teleportation \cite{teleportation}, quantum key distribution (QKD) \cite{qkd}, quantum secure direct communication (QSDC) \cite{qsdc1,qsdc2} and other important quantum communication protocols all require  entanglement to set up quantum channel. In a standard quantum communication protocol, one of the communication parties should create the entanglement locally and distribute it to another to share  entanglement. Unfortunately, during the entanglement distribution, the entanglement will suffer from the noise. The noise will make the photon loss and decoherence. If the photon is lost, the task  of communication is a failure. On the other hand, if the entanglement is decoherence, it will induce error, which will decrease the communication efficiency. More seriously, it will make the communication insecure.

Quantum repeaters \cite{repeater1} and quantum state amplification \cite{Ralph} are two powerful approaches to overcome the obstacle of photon loss. Quantum repeaters can connect the short distance entanglement to long distance entanglement. There are some important works of quantum repeaters in various
physical systems, such as nitrogen vacancy (NV) centers in diamond \cite{repeater2}, atomic ensembles \cite{repeater3}, and optical microcavities \cite{repeater4}. Quantum state amplification is used to increase the probability of the single photon after transmission. There are also some important theory and experiment works in quantum state amplification \cite{amplification1,amplification2,amplification3,amplification4,amplification5,amplification6,amplification7,amplification8}, such as the single qubit amplification \cite{amplification2}, the single qubit amplification encoded in polarization \cite{amplification3,amplification4}  and time-bin degrees of freedom \cite{amplification6}, single-photon entanglement amplification \cite{amplification7,amplification8}, and so on. On the other hand, entanglement concentration and entanglement purification are two important approaches to recover the degraded entangled states to maximally entangled states and high quality entangled states, respectively \cite {concentration1,concentration2,concentration3,concentration4,concentration5,concentration6,concentration7,concentration8}. Compared with entanglement concentration, entanglement purification which will be detailed  is a more general model, which focuses on the mixed states \cite{purification1,purification2,
purification3,purification5,purification6,purification7,purification8,purification9,purification10,purification11,purification12,purification13,purification14,
purification15,purification16,purification17,purification18,purification19,purification20,purification21,purification22,purification23,purification24}. In 1996, Bennett \emph{et al.} proposed the concept of entanglement purification \cite{purification1}. In 2001, Pan \emph{et al.} described a feasible entanglement purification protocol (EPP) in linear optics \cite{purification3}. Based on the cross-Kerr nonlinearity, Sheng \emph{et al.} presented an EPP with practical parametric down-conversion sources \cite{purification6}. Their protocol can be repeated to obtain a higher fidelity of mixed states. In 2010, the deterministic EPP were proposed with the success probability of 100\% in principle \cite{purification7}. In 2012, Ren \emph{et al.} described the first EPP model for hyperentanglement \cite{purification15}. Recent work shows that the EPP can be used to benefit the secure double-server blind quantum computation in a noise environment \cite{purification17}. There are also some other important EPPs for solid quantum systems, such as atoms \cite{purification18,purification19}, spins \cite{purification20,purification21,purification22}, and so on.

Logic-qubit entanglement  encodes many physical qubits in a logic qubits, which has been investigated  recently \cite{cghz1,cghz2,cghz3,cghz4,yan,pan,logicbell1,logicbell2,logicbell3,logicbell4,logicconcentration}. In 2011, Fr\"{o}wis and D\"{u}r first discussed a new type of logic-qubit entanglement, called the concatenated Greenberger-Horne-Zeiglinger (C-GHZ) state \cite{cghz1}. The C-GHZ state can be described as
\begin{eqnarray}
|\Phi_{1}^{\pm}\rangle_{N,m}=\frac{1}{\sqrt{2}}(|GHZ^{+}_{m}\rangle^{\otimes N} \pm |GHZ^{-}_{m}\rangle^{\otimes N}).\label{logic}
\end{eqnarray}
Here, $N$ is the number of logic qubit and $m$ is the number of physical qubit in each logic qubit, respectively.
States $|GHZ^{\pm}_{M}\rangle$ are the standard $m$-photon polarized GHZ states as
\begin{eqnarray}
|GHZ^{\pm}_{m}\rangle=\frac{1}{\sqrt{2}}(|H\rangle^{\otimes m}\pm|V\rangle^{\otimes m}).\label{phyghz}
\end{eqnarray}
Here $|H\rangle$ and $|V\rangle$ are the  horizonal polarized photon and  the vertical polarized photon, respectively.
In 2014, Lu \emph{et al.} reported the first experiment realization of C-GHZ state \cite{pan}. Recently, the logic-qubit entanglement generation \cite{yan},  analysis \cite{logicbell1,logicbell2,logicbell3,logicbell4} and entanglement concentration protocols \cite{logicconcentration} were proposed. Previous works of both theory and experiment show that it has the potential application in future quantum communication.

In this paper, we will investigate the entanglement distillation for  logic-qubit entanglement. For point to point quantum communication, we focus on the protocol for the two-logic-qubit entanglement. Previous EPPs cannot deal with logic-qubit entanglement, for they can only purify the errors in physical qubit, while logic-qubit entanglement not only contains the physical-qubit error, but also contains logic-qubit error. We mainly describe the distillation of three kinds of errors. The first is the logic bit-flip error. The second is the logic phase-flip error and the third is the physical bit-flip error. The physical phase-flip error equals to logic bit-flip error.  This paper is organized as follows. In Sec. II, we first explain the distillation of the logic bit-flip error. In Sec. III, we describe the distillation of the logic phase-flip error. In Sec. VI, we describe the correction of physical bit-flip error. In Sec. V, we extend this protocol to the case that the logic qubit is the arbitrary $m$-photon GHZ state. In Sec. VI, we present a discussion and a conclusion.

\section{Distillation of logic bit-flip error}
Now we start to explain the distillation protocol. The four two-logic-qubit entanglement can be described as
\begin{eqnarray}
|\Phi^{\pm}\rangle_{AB}=\frac{1}{\sqrt{2}}(|\phi^{+}\rangle_{A}|\phi^{+}\rangle_{B}\pm|\phi^{-}\rangle_{A}|\phi^{-}\rangle_{B}),\nonumber\\
|\Psi^{\pm}\rangle_{AB}=\frac{1}{\sqrt{2}}(|\phi^{+}\rangle_{A}|\phi^{-}\rangle_{B}\pm|\phi^{-}\rangle_{A}|\phi^{+}\rangle_{B}).\label{bell1}
\end{eqnarray}
States in Eq. (\ref{bell1}) essentially is the logic Bell states. Such states can be regarded as the  C-GHZ states with $m=N=2$.
Suppose two distant parties Alice and Bob want to share the maximally entangled state $|\Phi^{+}\rangle_{AB}$. Unfortunately, the noise will make the entanglement degrade. Generally, there are two kinds of error modes. The first error mode is the logic  error. It contains two errors. The first is the logic bit-flip error. It will make one of the logic qubit $|\phi^{+}\rangle$ become $|\phi^{-}\rangle$, and $|\phi^{-}\rangle$ become $|\phi^{+}\rangle$, respectively. If a logic qubit-flip error occurs with a probability of $1-F$, they will obtain a mixed state
\begin{eqnarray}
\rho_{LB}=F|\Phi^{+}\rangle\langle\Phi^{+}|+(1-F)|\Psi^{+}\rangle\langle\Psi^{+}|.\label{logicbit}
\end{eqnarray}
The second is the logic phase-flip error. It will make $|\Phi^{+}\rangle$ become $|\Phi^{-}\rangle$, and $|\Phi^{-}\rangle$ become $|\Phi^{+}\rangle$, respectively.  The second error mode is the physical error. It also contains two errors. The first is the physical bit-flip error and the second is
the physical phase-flip error. If  the first physical qubit in logic qubit $A$ suffers from the bit-flip error, it will make $|\phi^{+}\rangle_{A}$ become $|\psi^{+}\rangle_{A}$ and $|\psi^{+}\rangle_{A}$ become $|\phi^{+}\rangle_{A}$, respectively.  The whole logic-qubit entanglement $|\Phi^{+}\rangle_{AB}$ becomes
\begin{eqnarray}
|\Upsilon^{+}\rangle_{AB}=\frac{1}{\sqrt{2}}(|\psi^{+}\rangle_{A}|\phi^{+}\rangle_{B}+|\psi^{-}\rangle_{A}|\phi^{-}\rangle_{B}).
\end{eqnarray}
On the other hand, if the physical phase-flip error occurs on the first logic qubit, it will make $|\phi^{+}\rangle_{A}$ become $|\phi^{-}\rangle_{A}$ and $|\phi^{-}\rangle_{A}$ become $|\phi^{+}\rangle_{A}$, respectively. Interestingly, the physical phase-flip error essentially is the logic bit-flip error
as shown in Eq. (\ref{logicbit}). Therefore, the distillation task is to correct  three kinds of errors. They are logic bit-flip error, logic phase-flip error and physical bit-flip error.
\begin{figure}[!h]
\begin{center}
\includegraphics[width=9cm,angle=0]{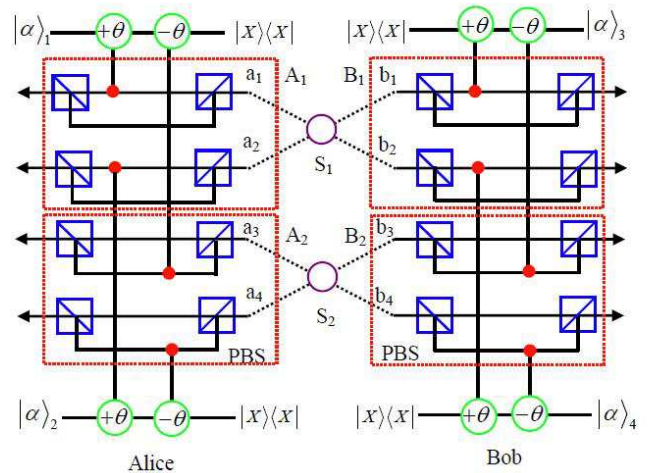}
\caption{A schematic drawing of our distillation protocol. S$_{1}$ and S$_{2}$ are two sources which prepare two copies of mixed states. $|\alpha\rangle_{1}$, $|\alpha\rangle_{2}$, $|\alpha\rangle_{3}$ and  $|\alpha\rangle_{4}$ are four coherent states.}
\end{center}
\end{figure}

Before we start to explain our protocol, we first briefly describe the quantum nondemolition (QND) measurement, which is the key element in this protocol.
 The cross-Kerr nonlinearity plays an important role in QND measurement.  As shown in Fig. 1, if an $|H\rangle$ polarization photon is in the $a_{1}$ spatial modes, it will pass through the polarization beam splitter (PBS). Here the PBS can transmit the $|H\rangle$ photon and reflect the $|V\rangle$ photon, respectively. For the $|H\rangle$  photon and the coherent state $|\alpha\rangle$, the cross-Kerr interaction causes the combined system to evolve as $|H\rangle|\alpha\rangle\rightarrow|H\rangle|\alpha e^{i\theta}\rangle$ \cite{cnot1,cnot2}. We find that the coherent state occurs a phase shift $\theta$. Here $\theta=\chi t$. $t$ is the interaction time and $\chi$ is the coupling strength of the nonlinearity, which is decided by the property of the nonlinear material. Therefore, by measuring the phase shift, we can judge the single photon, and do not measure the single photon itself. It is called the QND measurement. The QND based on the cross-Kerr nonlinearity has been widely used in quantum information processing \cite{cnot1,cnot2,guoqi,kerr1,kerr2,kerr3,kerr4,kerr5,kerr6}, such as  construction of the controlled-not gate \cite{cnot1,cnot2}, performing the  Bell-state analysis \cite{logicbell2}, entanglement purification \cite{purification6,purification7} and concentration \cite{concentration3,concentration4,concentration5}, preparing the entanglement state \cite{kerr3,kerr4,kerr5,kerr6}, and so on.

We first describe the distillation of the logic bit-flip error. As shown in Fig. 1, Alice and Bob share two copies of mixed logic-qubit entanglement $\rho_{A_{1}B_{1}}$ and $\rho_{A_{2}B_{2}}$, which has the same form in Eq. (\ref{logicbit}). State $\rho_{A_{1}B_{1}}$ is in the spatial modes $a_{1}$, $a_{2}$, $b_{1}$  and $b_{2}$, respectively. State
$\rho_{A_{2}B_{2}}$ is in the spatial modes $a_{3}$, $a_{4}$, $b_{3}$  and $b_{4}$, respectively. The whole system can be described as follows.
With the probability of $F^{2}$, it is in the state $|\Phi^{+}\rangle_{A_{1}B_{1}}|\Phi^{+}\rangle_{A_{2}B_{2}}$. With the equal probability $F(1-F)$, they are in the states $|\Phi^{+}\rangle_{A_{1}B_{1}}|\Psi^{+}\rangle_{A_{2}B_{2}}$ and $|\Psi^{+}\rangle_{A_{1}B_{1}}|\Phi^{+}\rangle_{A_{2}B_{2}}$, respectively.  With the probability of $(1-F)^{2}$, it is  in the state $|\Psi^{+}\rangle_{A_{1}B_{1}}|\Psi^{+}\rangle_{A_{2}B_{2}}$.

 We first discuss $|\Phi^{+}\rangle_{A_{1}B_{1}}|\Phi^{+}\rangle_{A_{2}B_{2}}$. Before the state passing through the PBSs, Alice and Bob both perform the Hadamard operation on
each photon. The quarter half-wave plate (QWP) can act as the role of Hadamard operation. The state can be described as
\begin{eqnarray}
&&|\Phi^{+}\rangle_{A_{1}B_{1}}|\Phi^{+}\rangle_{A_{2}B_{2}}\nonumber\\
&=&\frac{1}{\sqrt{2}}(|\phi^{+}\rangle_{A_{1}}|\phi^{+}\rangle_{B_{1}}+|\phi^{-}\rangle_{A_{1}}|\phi^{-}\rangle_{B_{1}})\nonumber\\
&\otimes&\frac{1}{\sqrt{2}}(|\phi^{+}\rangle_{A_{2}}|\phi^{+}\rangle_{B_{2}}+|\phi^{-}\rangle_{A_{2}}|\phi^{-}\rangle_{B_{2}})\nonumber\\
&\rightarrow&\frac{1}{\sqrt{2}}(|\phi^{+}\rangle_{A_{1}}|\phi^{+}\rangle_{B_{1}}+|\psi^{+}\rangle_{A_{1}}|\psi^{+}\rangle_{B_{1}})\nonumber\\
&\otimes&\frac{1}{\sqrt{2}}(|\phi^{+}\rangle_{A_{2}}|\phi^{+}\rangle_{B_{2}}+|\psi^{+}\rangle_{A_{2}}|\psi^{+}\rangle_{B_{2}})\nonumber\\
&=&\frac{1}{2}(|\phi^{+}\rangle_{A_{1}}|\phi^{+}\rangle_{A_{2}}|\phi^{+}\rangle_{B_{1}}|\phi^{+}\rangle_{B_{2}}\nonumber\\
&+&|\phi^{+}\rangle_{A_{1}}|\psi^{+}\rangle_{A_{2}}|\phi^{+}\rangle_{B_{1}}|\psi^{+}\rangle_{B_{2}}\nonumber\\
&+&|\psi^{+}\rangle_{A_{1}}|\phi^{+}\rangle_{A_{2}}|\psi^{+}\rangle_{B_{1}}|\phi^{+}\rangle_{B_{2}}\nonumber\\
&+&|\psi^{+}\rangle_{A_{1}}|\psi^{+}\rangle_{A_{2}}|\psi^{+}\rangle_{B_{1}}|\psi^{+}\rangle_{B_{2}}).\label{state1}
\end{eqnarray}

We discuss the first item $|\phi^{+}\rangle_{A_{1}}|\phi^{+}\rangle_{A_{2}}|\phi^{+}\rangle_{B_{1}}|\phi^{+}\rangle_{B_{2}}$.
Item $|\phi^{+}\rangle_{A_{1}}|\phi^{+}\rangle_{A_{2}}|\phi^{+}\rangle_{B_{1}}|\phi^{+}\rangle_{B_{2}}$ combined with the cross-Kerr nonlinearity can be evolved as
\begin{eqnarray}
&&|\phi^{+}\rangle_{A_{1}}|\phi^{+}\rangle_{A_{2}}|\phi^{+}\rangle_{B_{1}}|\phi^{+}\rangle_{B_{2}}
|\alpha\rangle_{1}|\alpha\rangle_{2}|\alpha\rangle_{3}|\alpha\rangle_{4}\nonumber\\
&=&\frac{1}{\sqrt{2}}(|H\rangle_{a_{1}}|H\rangle_{a_{2}}+|V\rangle_{a_{1}}|V\rangle_{a_{2}})\frac{1}{\sqrt{2}}(|H\rangle_{a_{3}}|H\rangle_{a_{4}}\nonumber\\
&+&|V\rangle_{a_{3}}|V\rangle_{a_{4}})|\alpha\rangle_{1}|\alpha\rangle_{2}\frac{1}{\sqrt{2}}(|H\rangle_{b_{1}}|H\rangle_{b_{2}}+|V\rangle_{b_{1}}|V\rangle_{b_{2}})\nonumber\\
&&\frac{1}{\sqrt{2}}(|H\rangle_{b_{3}}|H\rangle_{b_{4}}+|V\rangle_{b_{3}}|V\rangle_{b_{4}})|\alpha\rangle_{3}|\alpha\rangle_{4}\nonumber\\
&\rightarrow&\frac{1}{2}(|H\rangle_{a_{1}}|H\rangle_{a_{2}}|H\rangle_{a_{3}}|H\rangle_{a_{4}}|\alpha e^{i\theta}\rangle_{1}|\alpha e^{i\theta}\rangle_{2}\nonumber\\
&+&|H\rangle_{a_{1}}|H\rangle_{a_{2}}|V\rangle_{a_{3}}|V\rangle_{a_{4}}|\alpha\rangle_{1}|\alpha\rangle_{2}\nonumber\\
&+&|V\rangle_{a_{1}}|V\rangle_{a_{2}}|H\rangle_{a_{3}}|H\rangle_{a_{4}}|\alpha\rangle_{1}|\alpha\rangle_{2}\nonumber\\
&+&|V\rangle_{a_{1}}|V\rangle_{a_{2}}|V\rangle_{a_{3}}|V\rangle_{a_{4}}|\alpha e^{-i\theta}\rangle_{1}|\alpha e^{-i\theta}\rangle_{2})\nonumber\\
&\otimes&\frac{1}{2}(|H\rangle_{b_{1}}|H\rangle_{b_{2}}|H\rangle_{b_{3}}|H\rangle_{b_{4}}|\alpha e^{i\theta}\rangle_{3}|\alpha e^{i\theta}\rangle_{4}\nonumber\\
&+&|H\rangle_{b_{1}}|H\rangle_{b_{2}}|V\rangle_{b_{3}}|V\rangle_{b_{4}}|\alpha\rangle_{3}|\alpha\rangle_{4}\nonumber\\
&+&|V\rangle_{b_{1}}|V\rangle_{b_{2}}|H\rangle_{b_{3}}|H\rangle_{b_{4}}|\alpha\rangle_{3}|\alpha\rangle_{4}\nonumber\\
&+&|V\rangle_{b_{1}}|V\rangle_{b_{2}}|V\rangle_{b_{3}}|V\rangle_{b_{4}}|\alpha e^{-i\theta}\rangle_{3}|\alpha e^{-i\theta}\rangle_{4}).\label{evolve1}
\end{eqnarray}
They pick up the cases that all the coherent states make a phase shift $\theta$. Here we exploit the $|X\rangle\langle X|$ homodyne detection, which can make
$\pm \theta$ do not distinguish \cite{cnot1}. Such selection condition will make the state in Eq. (\ref{evolve1}) become
\begin{eqnarray}
&\rightarrow&(|H\rangle_{a_{1}}|H\rangle_{a_{2}}|H\rangle_{a_{3}}|H\rangle_{a_{4}}+|V\rangle_{a_{1}}|V\rangle_{a_{2}}|V\rangle_{a_{3}}|V\rangle_{a_{4}})\nonumber\\
&\otimes&(|H\rangle_{b_{1}}|H\rangle_{b_{2}}|H\rangle_{b_{3}}|H\rangle_{b_{4}}+|V\rangle_{b_{1}}|V\rangle_{b_{2}}|V\rangle_{b_{3}}|V\rangle_{b_{4}}).\label{collapse1}\nonumber\\
\end{eqnarray}
Items $|\phi^{+}\rangle_{A_{1}}|\psi^{+}\rangle_{A_{2}}|\phi^{+}\rangle_{B_{1}}|\psi^{+}\rangle_{B_{2}}$ and $|\psi^{+}\rangle_{A_{1}}|\phi^{+}\rangle_{A_{2}}|\psi^{+}\rangle_{B_{1}}|\phi^{+}\rangle_{B_{2}}$ cannot make all the coherent states pick up a phase shift. For example, item $|\phi^{+}\rangle_{A_{1}}|\psi^{+}\rangle_{A_{2}}|\phi^{+}\rangle_{B_{1}}|\psi^{+}\rangle_{B_{2}}$ combined with the coherent states can be evolved as
\begin{eqnarray}
&&|\phi^{+}\rangle_{A_{1}}|\psi^{+}\rangle_{A_{2}}|\phi^{+}\rangle_{B_{1}}|\psi^{+}\rangle_{B_{2}}|\alpha\rangle_{1}|\alpha\rangle_{2}|\alpha\rangle_{3}|\alpha\rangle_{4}\nonumber\\
&=&\frac{1}{\sqrt{2}}(|H\rangle_{a_{1}}|H\rangle_{a_{2}}+|V\rangle_{a_{1}}|V\rangle_{a_{2}})\frac{1}{\sqrt{2}}(|H\rangle_{a_{3}}|V\rangle_{a_{4}}\nonumber\\
&+&|V\rangle_{a_{3}}|H\rangle_{a_{4}})|\alpha\rangle_{1}|\alpha\rangle_{2}\frac{1}{\sqrt{2}}(|H\rangle_{b_{1}}|H\rangle_{b_{2}}+|V\rangle_{b_{1}}|V\rangle_{b_{2}})\nonumber\\
&&\frac{1}{\sqrt{2}}(|H\rangle_{b_{3}}|V\rangle_{b_{4}}+|V\rangle_{b_{3}}|H\rangle_{b_{4}})|\alpha\rangle_{3}|\alpha\rangle_{4}\nonumber\\
&\rightarrow&\frac{1}{2}(|H\rangle_{a_{1}}|H\rangle_{a_{2}}|H\rangle_{a_{3}}|V\rangle_{a_{4}}|\alpha e^{i\theta}\rangle_{1}|\alpha\rangle_{2}\nonumber\\
&+&|H\rangle_{a_{1}}|H\rangle_{a_{2}}|V\rangle_{a_{3}}|H\rangle_{a_{4}}|\alpha\rangle_{1}|\alpha e^{i\theta}\rangle_{2}\nonumber\\
&+&|V\rangle_{a_{1}}|V\rangle_{a_{2}}|H\rangle_{a_{3}}|V\rangle_{a_{4}}|\alpha\rangle_{1}|\alpha e^{-i\theta}\rangle_{2}\nonumber\\
&+&|V\rangle_{a_{1}}|V\rangle_{a_{2}}|V\rangle_{a_{3}}|H\rangle_{a_{4}}|\alpha e^{-i\theta}\rangle_{1}|\alpha\rangle_{2})\nonumber\\
&\otimes&\frac{1}{2}(|H\rangle_{b_{1}}|H\rangle_{b_{2}}|H\rangle_{b_{3}}|V\rangle_{b_{4}}|\alpha e^{i\theta}\rangle_{3}|\alpha\rangle_{4}\nonumber\\
&+&|H\rangle_{b_{1}}|H\rangle_{b_{2}}|V\rangle_{b_{3}}|H\rangle_{b_{4}}|\alpha\rangle_{3}|\alpha e^{i\theta}\rangle_{4}\nonumber\\
&+&|V\rangle_{b_{1}}|V\rangle_{b_{2}}|H\rangle_{b_{3}}|V\rangle_{b_{4}}|\alpha\rangle_{3}|\alpha e^{-i\theta}\rangle_{4}\nonumber\\
&+&|V\rangle_{b_{1}}|V\rangle_{b_{2}}|V\rangle_{b_{3}}|H\rangle_{b_{4}}|\alpha e^{-i\theta}\rangle_{3}|\alpha\rangle_{4}).\label{evolve2}
\end{eqnarray}
 Finally, let us discuss the last item
$|\psi^{+}\rangle_{A_{1}}|\psi^{+}\rangle_{A_{2}}|\psi^{+}\rangle_{B_{1}}|\psi^{+}\rangle_{B_{2}}$. It combined with the coherent states can be evolved as
\begin{eqnarray}
&&|\psi^{+}\rangle_{A_{1}}|\psi^{+}\rangle_{A_{2}}|\psi^{+}\rangle_{B_{1}}|\psi^{+}\rangle_{B_{2}}|\alpha\rangle_{1}|\alpha\rangle_{2}|\alpha\rangle_{3}|\alpha\rangle_{4}\nonumber\\
&=&\frac{1}{\sqrt{2}}(|H\rangle_{a_{1}}|V\rangle_{a_{2}}+|V\rangle_{a_{1}}|H\rangle_{a_{2}})\frac{1}{\sqrt{2}}(|H\rangle_{a_{3}}|V\rangle_{a_{4}}\nonumber\\
&+&|V\rangle_{a_{3}}|H\rangle_{a_{4}})|\alpha\rangle_{1}|\alpha\rangle_{2}\frac{1}{\sqrt{2}}(|H\rangle_{b_{1}}|V\rangle_{b_{2}}+|V\rangle_{b_{1}}|H\rangle_{b_{2}})\nonumber\\
&&\frac{1}{\sqrt{2}}(|H\rangle_{b_{3}}|V\rangle_{b_{4}}+|V\rangle_{b_{3}}|H\rangle_{b_{4}})|\alpha\rangle_{3}|\alpha\rangle_{4}\nonumber\\
&\rightarrow&\frac{1}{2}(|H\rangle_{a_{1}}|V\rangle_{a_{2}}|H\rangle_{a_{3}}|V\rangle_{a_{4}}|\alpha e^{i\theta}\rangle_{1}|\alpha e^{-i\theta}\rangle_{2}\nonumber\\
&+&|H\rangle_{a_{1}}|V\rangle_{a_{2}}|V\rangle_{a_{3}}|H\rangle_{a_{4}}|\alpha\rangle_{1}|\alpha\rangle_{2}\nonumber\\
&+&|V\rangle_{a_{1}}|H\rangle_{a_{2}}H\rangle_{a_{3}}|V\rangle_{a_{4}}|\alpha\rangle_{1}|\alpha\rangle_{2}\nonumber\\
&+&|V\rangle_{a_{1}}|H\rangle_{a_{2}}|V\rangle_{a_{3}}|H\rangle_{a_{4}}|\alpha e^{-i\theta}\rangle_{1}|\alpha e^{i\theta}\rangle_{2})\nonumber\\
&\otimes&\frac{1}{2}(|H\rangle_{b_{1}}|V\rangle_{b_{2}}|H\rangle_{b_{3}}|V\rangle_{b_{4}}|\alpha e^{i\theta}\rangle_{3}|\alpha e^{-i\theta}\rangle_{4}\nonumber\\
&+&|H\rangle_{b_{1}}|V\rangle_{b_{2}}|V\rangle_{b_{3}}|H\rangle_{b_{4}}|\alpha\rangle_{3}|\alpha\rangle_{4}\nonumber\\
&+&|V\rangle_{b_{1}}|H\rangle_{b_{2}}H\rangle_{b_{3}}|V\rangle_{b_{4}}|\alpha\rangle_{3}|\alpha\rangle_{4}\nonumber\\
&+&|V\rangle_{b_{1}}|H\rangle_{b_{2}}|V\rangle_{b_{3}}|H\rangle_{b_{4}}|\alpha e^{-i\theta}\rangle_{3}|\alpha e^{i\theta}\rangle_{4}).\label{evolve3}
\end{eqnarray}
Therefore, by picking up the case that all the coherent states make a phase shift $\theta$, they will obtain
\begin{eqnarray}
&&\frac{1}{4}[(|H\rangle_{a_{1}}|H\rangle_{a_{2}}|H\rangle_{a_{3}}|H\rangle_{a_{4}}+|V\rangle_{a_{1}}|V\rangle_{a_{2}}|V\rangle_{a_{3}}|V\rangle_{a_{4}})\nonumber\\
&\otimes&(|H\rangle_{b_{1}}|H\rangle_{b_{2}}|H\rangle_{b_{3}}|H\rangle_{b_{4}}+|V\rangle_{b_{1}}|V\rangle_{b_{2}}|V\rangle_{b_{3}}|V\rangle_{b_{4}})\nonumber\\
&+&(|H\rangle_{a_{1}}|V\rangle_{a_{2}}|H\rangle_{a_{3}}|V\rangle_{a_{4}}+|V\rangle_{a_{1}}|H\rangle_{a_{2}}|V\rangle_{a_{3}}|H\rangle_{a_{4}})\nonumber\\
&\otimes&(|H\rangle_{b_{1}}|V\rangle_{b_{2}}|H\rangle_{b_{3}}|V\rangle_{b_{4}}+|V\rangle_{b_{1}}|H\rangle_{b_{2}}|V\rangle_{b_{3}}|H\rangle_{b_{4}})].\label{result1}\nonumber\\
\end{eqnarray}

Interestingly, state $|\Phi^{+}\rangle_{A_{1}B_{1}}|\Psi^{+}\rangle_{A_{2}B_{2}}$ and $|\Psi^{+}\rangle_{A_{1}B_{1}}|\Phi^{+}\rangle_{A_{2}B_{2}}$ cannot lead the case that all the coherent states pick up the phase shift $\theta$. For example, we take state $|\Phi^{+}\rangle_{A_{1}B_{1}}|\Psi^{+}\rangle_{A_{2}B_{2}}$ for example. By perform the Hadamard operation on all the photons, state $|\Phi^{+}\rangle_{A_{1}B_{1}}|\Psi^{+}\rangle_{A_{2}B_{2}}$ can be written as
\begin{eqnarray}
&&|\Phi^{+}\rangle_{A_{1}B_{1}}|\Psi^{+}\rangle_{A_{2}B_{2}}\nonumber\\
&=&\frac{1}{\sqrt{2}}(|\phi^{+}\rangle_{A_{1}}|\phi^{+}\rangle_{B_{1}}+|\phi^{-}\rangle_{A_{1}}|\phi^{-}\rangle_{B_{1}})\nonumber\\
&\otimes&\frac{1}{\sqrt{2}}(|\phi^{+}\rangle_{A_{2}}|\phi^{-}\rangle_{B_{2}}+|\phi^{-}\rangle_{A_{2}}|\phi^{+}\rangle_{B_{2}})\nonumber\\
&\rightarrow&\frac{1}{\sqrt{2}}(|\phi^{+}\rangle_{A_{1}}|\phi^{+}\rangle_{B_{1}}+|\psi^{+}\rangle_{A_{1}}|\psi^{+}\rangle_{B_{1}})\nonumber\\
&\otimes&\frac{1}{\sqrt{2}}(|\phi^{+}\rangle_{A_{2}}|\psi^{+}\rangle_{B_{2}}+|\psi^{+}\rangle_{A_{2}}|\phi^{+}\rangle_{B_{2}})\nonumber\\
&=&\frac{1}{2}(|\phi^{+}\rangle_{A_{1}}|\phi^{+}\rangle_{A_{2}}|\phi^{+}\rangle_{B_{1}}|\psi^{+}\rangle_{B_{2}}\nonumber\\
&+&|\phi^{+}\rangle_{A_{1}}|\psi^{+}\rangle_{A_{2}}|\phi^{+}\rangle_{B_{1}}|\phi^{+}\rangle_{B_{2}}\nonumber\\
&+&|\psi^{+}\rangle_{A_{1}}|\phi^{+}\rangle_{A_{2}}|\psi^{+}\rangle_{B_{1}}|\psi^{+}\rangle_{B_{2}}\nonumber\\
&+&|\psi^{+}\rangle_{A_{1}}|\psi^{+}\rangle_{A_{2}}|\psi^{+}\rangle_{B_{1}}|\phi^{+}\rangle_{B_{2}}).\label{state2}
\end{eqnarray}
All the items in Eq.(\ref{state2}) cannot lead the phase shift $\theta$. For example, the first item $|\phi^{+}\rangle_{A_{1}}|\phi^{+}\rangle_{A_{2}}|\phi^{+}\rangle_{B_{1}}|\psi^{+}\rangle_{B_{2}}$ combined with the coherent states can be evolved as
\begin{eqnarray}
&&|\phi^{+}\rangle_{A_{1}}|\phi^{+}\rangle_{A_{2}}|\phi^{+}\rangle_{B_{1}}|\psi^{+}\rangle_{B_{2}}|\alpha\rangle_{1}|\alpha\rangle_{2}|\alpha\rangle_{3}|\alpha\rangle_{4}\nonumber\\
&=&\frac{1}{\sqrt{2}}(|H\rangle_{a_{1}}|H\rangle_{a_{2}}+|V\rangle_{a_{1}}|V\rangle_{a_{2}})\frac{1}{\sqrt{2}}(|H\rangle_{a_{3}}|H\rangle_{a_{4}}\nonumber\\
&+&|V\rangle_{a_{3}}|V\rangle_{a_{4}})|\alpha\rangle_{1}|\alpha\rangle_{2}\frac{1}{\sqrt{2}}(|H\rangle_{b_{1}}|H\rangle_{b_{2}}+|V\rangle_{b_{1}}|V\rangle_{b_{2}})\nonumber\\
&&\frac{1}{\sqrt{2}}(|H\rangle_{b_{3}}|V\rangle_{b_{4}}+|V\rangle_{b_{3}}|H\rangle_{b_{4}})|\alpha\rangle_{3}|\alpha\rangle_{4}\nonumber\\
&\rightarrow&\frac{1}{2}(|H\rangle_{a_{1}}|H\rangle_{a_{2}}|H\rangle_{a_{3}}|H\rangle_{a_{4}}|\alpha e^{i\theta}\rangle_{1}|\alpha e^{i\theta}\rangle_{2}\nonumber\\
&+&|H\rangle_{a_{1}}|H\rangle_{a_{2}}|V\rangle_{a_{3}}|V\rangle_{a_{4}}|\alpha\rangle_{1}|\alpha\rangle_{2}\nonumber\\
&+&|V\rangle_{a_{1}}|V\rangle_{a_{2}}|H\rangle_{a_{3}}|H\rangle_{a_{4}}|\alpha\rangle_{1}|\alpha\rangle_{2}\nonumber\\
&+&|V\rangle_{a_{1}}|V\rangle_{a_{2}}|V\rangle_{a_{3}}|V\rangle_{a_{4}}|\alpha e^{-i\theta}\rangle_{1}|\alpha e^{-i\theta}\rangle_{2})\nonumber\\
&\otimes&\frac{1}{2}(|H\rangle_{b_{1}}|H\rangle_{b_{2}}|H\rangle_{b_{3}}|V\rangle_{b_{4}}|\alpha e^{i\theta}\rangle_{3}|\alpha\rangle_{4}\nonumber\\
&+&|H\rangle_{b_{1}}|H\rangle_{b_{2}}|V\rangle_{b_{3}}|H\rangle_{b_{4}}|\alpha\rangle_{3}|\alpha e^{i\theta}\rangle_{4}\nonumber\\
&+&|V\rangle_{b_{1}}|V\rangle_{b_{2}}|H\rangle_{b_{3}}|V\rangle_{b_{4}}|\alpha e^{-i\theta}\rangle_{3}|\alpha\rangle_{4}\nonumber\\
&+&|V\rangle_{b_{1}}|V\rangle_{b_{2}}|V\rangle_{b_{3}}|H\rangle_{b_{4}}|\alpha e^{-i\theta}\rangle_{3}|\alpha\rangle_{4}).\label{evolve4}
\end{eqnarray}
As shown in Eq. (\ref{evolve4}), all the possible cases cannot satisfy the condition that all the coherent states pick up the phase shift $\theta$. Therefore, it can be excluded automatically. Certainly, the other items such as $|\phi^{+}\rangle_{A_{1}}|\psi^{+}\rangle_{A_{2}}|\phi^{+}\rangle_{B_{1}}|\phi^{+}\rangle_{B_{2}}$, $|\psi^{+}\rangle_{A_{1}}|\phi^{+}\rangle_{A_{2}}|\psi^{+}\rangle_{B_{1}}|\psi^{+}\rangle_{B_{2}}$ and $|\psi^{+}\rangle_{A_{1}}|\psi^{+}\rangle_{A_{2}}|\psi^{+}\rangle_{B_{1}}|\phi^{+}\rangle_{B_{2}}$ also cannot satisfy the the condition that all the coherent states pick up the phase shift $\theta$. On the other hand, with the same principle, state $|\Psi^{+}\rangle_{A_{1}B_{1}}|\Phi^{+}\rangle_{A_{2}B_{2}}$ also cannot lead all the coherent states pick up the phase shift $\theta$ and can be excluded automatically.

Finally, let us discuss the state $|\Psi^{+}\rangle_{A_{1}B_{1}}|\Psi^{+}\rangle_{A_{2}B_{2}}$. After performing the Hadamard operations on all the photons,
it becomes
\begin{eqnarray}
&&|\Psi^{+}\rangle_{A_{1}B_{1}}|\Psi^{+}\rangle_{A_{2}B_{2}}\nonumber\\
&=&\frac{1}{\sqrt{2}}(|\phi^{+}\rangle_{A_{1}}|\phi^{-}\rangle_{B_{1}}+|\phi^{-}\rangle_{A_{1}}|\phi^{+}\rangle_{B_{1}})\nonumber\\
&\otimes&\frac{1}{\sqrt{2}}(|\phi^{+}\rangle_{A_{2}}|\phi^{-}\rangle_{B_{2}}+|\phi^{-}\rangle_{A_{2}}|\phi^{+}\rangle_{B_{2}})\nonumber\\
&\rightarrow&\frac{1}{\sqrt{2}}(|\phi^{+}\rangle_{A_{1}}|\psi^{+}\rangle_{B_{1}}+|\psi^{+}\rangle_{A_{1}}|\phi^{+}\rangle_{B_{1}})\nonumber\\
&\otimes&\frac{1}{\sqrt{2}}(|\phi^{+}\rangle_{A_{2}}|\psi^{+}\rangle_{B_{2}}+|\psi^{+}\rangle_{A_{2}}|\phi^{+}\rangle_{B_{2}})\nonumber\\
&=&\frac{1}{2}(|\phi^{+}\rangle_{A_{1}}|\phi^{+}\rangle_{A_{2}}|\psi^{+}\rangle_{B_{1}}|\psi^{+}\rangle_{B_{2}}\nonumber\\
&+&|\phi^{+}\rangle_{A_{1}}|\psi^{+}\rangle_{A_{2}}|\psi^{+}\rangle_{B_{1}}|\phi^{+}\rangle_{B_{2}}\nonumber\\
&+&|\psi^{+}\rangle_{A_{1}}|\phi^{+}\rangle_{A_{2}}|\phi^{+}\rangle_{B_{1}}|\psi^{+}\rangle_{B_{2}}\nonumber\\
&+&|\psi^{+}\rangle_{A_{1}}|\psi^{+}\rangle_{A_{2}}|\phi^{+}\rangle_{B_{1}}|\phi^{+}\rangle_{B_{2}}).\label{state3}
\end{eqnarray}
The first item $|\phi^{+}\rangle_{A_{1}}|\phi^{+}\rangle_{A_{2}}|\psi^{+}\rangle_{B_{1}}|\psi^{+}\rangle_{B_{2}}$ combined with four coherent states can be evolved as
\begin{eqnarray}
&&|\phi^{+}\rangle_{A_{1}}|\phi^{+}\rangle_{A_{2}}|\psi^{+}\rangle_{B_{1}}|\psi^{+}\rangle_{B_{2}}|\alpha\rangle_{1}|\alpha\rangle_{2}|\alpha\rangle_{3}|\alpha\rangle_{4}\nonumber\\
&=&\frac{1}{\sqrt{2}}(|H\rangle_{a_{1}}|H\rangle_{a_{2}}+|V\rangle_{a_{1}}|V\rangle_{a_{2}})\frac{1}{\sqrt{2}}(|H\rangle_{a_{3}}|H\rangle_{a_{4}}\nonumber\\
&+&|V\rangle_{a_{3}}|V\rangle_{a_{4}})|\alpha\rangle_{1}|\alpha\rangle_{2}\frac{1}{\sqrt{2}}(|H\rangle_{b_{1}}|V\rangle_{b_{2}}+|V\rangle_{b_{1}}|H\rangle_{b_{2}})\nonumber\\
&&\frac{1}{\sqrt{2}}(|H\rangle_{b_{3}}|V\rangle_{b_{4}}+|V\rangle_{b_{3}}|H\rangle_{b_{4}})|\alpha\rangle_{3}|\alpha\rangle_{4}\nonumber\\
&&\frac{1}{\sqrt{2}}(|H\rangle_{b_{3}}|V\rangle_{b_{4}}+|V\rangle_{b_{3}}|H\rangle_{b_{4}})|\alpha\rangle_{3}|\alpha\rangle_{4}\nonumber\\
&\rightarrow&\frac{1}{2}(|H\rangle_{a_{1}}|H\rangle_{a_{2}}|H\rangle_{a_{3}}|H\rangle_{a_{4}}|\alpha e^{i\theta}\rangle_{1}|\alpha e^{i\theta}\rangle_{2}\nonumber\\
&+&|H\rangle_{a_{1}}|H\rangle_{a_{2}}|V\rangle_{a_{3}}|V\rangle_{a_{4}}|\alpha\rangle_{1}|\alpha\rangle_{2}\nonumber\\
&+&|V\rangle_{a_{1}}|V\rangle_{a_{2}}|H\rangle_{a_{3}}|H\rangle_{a_{4}}|\alpha\rangle_{1}|\alpha\rangle_{2}\nonumber\\
&+&|V\rangle_{a_{1}}|V\rangle_{a_{2}}|V\rangle_{a_{3}}|V\rangle_{a_{4}}|\alpha e^{-i\theta}\rangle_{1}|\alpha e^{-i\theta}\rangle_{2})\nonumber\\
&\otimes&\frac{1}{2}(|H\rangle_{b_{1}}|V\rangle_{b_{2}}|H\rangle_{b_{3}}|V\rangle_{b_{4}}|\alpha e^{i\theta}\rangle_{3}|\alpha e^{-i\theta}\rangle_{4}\nonumber\\
&+&|H\rangle_{b_{1}}|V\rangle_{b_{2}}|V\rangle_{b_{3}}|H\rangle_{b_{4}}|\alpha\rangle_{3}|\alpha \rangle_{4}\nonumber\\
&+&|V\rangle_{b_{1}}|H\rangle_{b_{2}}|H\rangle_{b_{3}}|V\rangle_{b_{4}}|\alpha\rangle_{3}|\alpha\rangle_{4}\nonumber\\
&+&|V\rangle_{b_{1}}|H\rangle_{b_{2}}|V\rangle_{b_{3}}|H\rangle_{b_{4}}|\alpha e^{-i\theta}\rangle_{3}|\alpha e^{i\theta}\rangle_{4}).\label{evolve5}
\end{eqnarray}
Similarly, items $|\phi^{+}\rangle_{A_{1}}|\psi^{+}\rangle_{A_{2}}|\psi^{+}\rangle_{B_{1}}|\phi^{+}\rangle_{B_{2}}$ and $|\psi^{+}\rangle_{A_{1}}|\phi^{+}\rangle_{A_{2}}|\phi^{+}\rangle_{B_{1}}|\psi^{+}\rangle_{B_{2}}$ cannot lead all the coherent states pick up the same phase shift $\theta$. The last item $|\psi^{+}\rangle_{A_{1}}|\psi^{+}\rangle_{A_{2}}|\phi^{+}\rangle_{B_{1}}|\psi^{+}\rangle_{B_{2}}$ can lead all the coherent states pick up the same phase shift. It combined with four coherent states can be evolved as
\begin{eqnarray}
&&|\psi^{+}\rangle_{A_{1}}|\psi^{+}\rangle_{A_{2}}|\phi^{+}\rangle_{B_{1}}|\phi^{+}\rangle_{B_{2}}|\alpha\rangle_{1}|\alpha\rangle_{2}|\alpha\rangle_{3}|\alpha\rangle_{4}\nonumber\\
&=&\frac{1}{\sqrt{2}}(|H\rangle_{a_{1}}|V\rangle_{a_{2}}+|V\rangle_{a_{1}}|H\rangle_{a_{2}})\frac{1}{\sqrt{2}}(|H\rangle_{a_{3}}|V\rangle_{a_{4}}\nonumber\\
&+&|V\rangle_{a_{3}}|H\rangle_{a_{4}})|\alpha\rangle_{1}|\alpha\rangle_{2}\frac{1}{\sqrt{2}}(|H\rangle_{b_{1}}|H\rangle_{b_{2}}+|V\rangle_{b_{1}}|V\rangle_{b_{2}})\nonumber\\
&&\frac{1}{\sqrt{2}}(|H\rangle_{b_{3}}|H\rangle_{b_{4}}+|V\rangle_{b_{3}}|V\rangle_{b_{4}})|\alpha\rangle_{3}|\alpha\rangle_{4}\nonumber\\
&\rightarrow&\frac{1}{2}(|H\rangle_{a_{1}}|V\rangle_{a_{2}}|H\rangle_{a_{3}}|V\rangle_{a_{4}}|\alpha e^{i\theta}\rangle_{3}|\alpha e^{-i\theta}\rangle_{4}\nonumber\\
&+&|H\rangle_{a_{1}}|V\rangle_{a_{2}}|V\rangle_{a_{3}}|H\rangle_{a_{4}}|\alpha\rangle_{3}|\alpha \rangle_{4}\nonumber\\
&+&|V\rangle_{a_{1}}|H\rangle_{a_{2}}|H\rangle_{a_{3}}|V\rangle_{a_{4}}|\alpha\rangle_{3}|\alpha\rangle_{4}\nonumber\\
&+&|V\rangle_{a_{1}}|H\rangle_{a_{2}}|V\rangle_{a_{3}}|H\rangle_{a_{4}}|\alpha e^{-i\theta}\rangle_{3}|\alpha e^{i\theta}\rangle_{4})\nonumber\\
&\otimes&\frac{1}{2}(|H\rangle_{b_{1}}|H\rangle_{b_{2}}|H\rangle_{b_{3}}|H\rangle_{b_{4}}|\alpha e^{i\theta}\rangle_{1}|\alpha e^{i\theta}\rangle_{2}\nonumber\\
&+&|H\rangle_{b_{1}}|H\rangle_{b_{2}}|V\rangle_{b_{3}}|V\rangle_{b_{4}}|\alpha\rangle_{1}|\alpha\rangle_{2}\nonumber\\
&+&|V\rangle_{b_{1}}|V\rangle_{b_{2}}|H\rangle_{b_{3}}|H\rangle_{b_{4}}|\alpha\rangle_{1}|\alpha\rangle_{2}\nonumber\\
&+&|V\rangle_{b_{1}}|V\rangle_{b_{2}}|V\rangle_{b_{3}}|V\rangle_{b_{4}}|\alpha e^{-i\theta}\rangle_{1}|\alpha e^{-i\theta}\rangle_{2}).\label{evolve6}
\end{eqnarray}
From Eqs. (\ref{evolve5}) and (\ref{evolve6}), if Alice and Bob pick up the case that all the coherent states show the same phase shift $\theta$, they will obtain
\begin{eqnarray}
&&\frac{1}{4}[(|H\rangle_{a_{1}}|H\rangle_{a_{2}}|H\rangle_{a_{3}}|H\rangle_{a_{4}}+|V\rangle_{a_{1}}|V\rangle_{a_{2}}|V\rangle_{a_{3}}|V\rangle_{a_{4}})\nonumber\\
&\otimes&(|H\rangle_{b_{1}}|V\rangle_{b_{2}}|H\rangle_{b_{3}}|V\rangle_{b_{4}}+|V\rangle_{b_{1}}|H\rangle_{b_{2}}|V\rangle_{b_{3}}|H\rangle_{b_{4}})\nonumber\\
&+&(|H\rangle_{a_{1}}|V\rangle_{a_{2}}|H\rangle_{a_{3}}|V\rangle_{a_{4}}+|V\rangle_{a_{1}}|H\rangle_{a_{2}}|V\rangle_{a_{3}}|H\rangle_{a_{4}})\nonumber\\
&\otimes&(|H\rangle_{b_{1}}|H\rangle_{b_{2}}|H\rangle_{b_{3}}|H\rangle_{b_{4}}+|V\rangle_{b_{1}}|V\rangle_{b_{2}}|V\rangle_{b_{3}}|V\rangle_{b_{4}})]\label{result2}.\nonumber\\
\end{eqnarray}
From above description, by selecting the case that all the coherent states pick up the same phase shift $\theta$, Alice and Bob will ultimately obtain the state in Eq. (\ref{result1}), with the probability of $\frac{F^{2}}{8}$, and obtain the state in Eq. (\ref{result2}), with the probability of $\frac{(1-F)^{2}}{8}$.
Finally, they measure the photons in the spatial modes $a_{3}$, $a_{4}$, $b_{3}$ and $b_{4}$ in $|\pm\rangle$ basis, respectively. Here $|\pm\rangle=\frac{1}{\sqrt{2}}(|H\rangle\pm|V\rangle)$.
 If the measurement result is are all the same, say  $|+\rangle_{a_{3}}|+\rangle_{a_{4}}|+\rangle_{b_{3}}|+\rangle_{b_{4}}$ or
 $|-\rangle_{a_{3}}|-\rangle_{a_{4}}|-\rangle_{b_{3}}|-\rangle_{b_{4}}$, they will obtain a new mixed state $\rho'_{A_{1}B_{1}}$
 \begin{eqnarray}
\rho'_{A_{1}B_{1}}=F'|\Phi^{+}\rangle_{A_{1}B_{1}}\langle\Phi^{+}|+(1-F')|\Psi^{+}\rangle_{A_{1}B_{1}}\langle\Psi^{+}|,\label{new1}
\end{eqnarray}
after performing the Hadamard operations on the photons in spatial modes $a_{1}$, $a_{2}$, $b_{1}$ and $b_{2}$.
 Here $F'=\frac{F^{2}}{F^{2}+(1-F)^{2}}$. If $F> \frac{1}{2}$, $F'>F$. On the other hand, if the measurement result in spatial modes  $a_{3}a_{4}$  are opposite, i. e. $|+\rangle_{a_{3}}|-\rangle_{a_{4}}$, or $|-\rangle_{a_{3}}|+\rangle_{a_{4}}$, they should perform a phase flip operation on one of the photons in $a_{1}$ or $a_{2}$ modes. If the measurement result in spatial modes $b_{3}b_{4}$ are opposite, i. e. $|+\rangle_{b_{3}}|-\rangle_{b_{4}}$, or $|-\rangle_{b_{3}}|+\rangle_{b_{4}}$, they should also perform a phase flip operation on one of the photons in $b_{1}$ or $b_{2}$ modes. In this way, they can obtain the same mixed state in Eq. (\ref{new1}).

 \section{Distillation of logic phase-flip error}
  The distillation of logic phase-flip error can be described as follows. Suppose Alice and Bob share the mixed state as
 \begin{eqnarray}
\rho_{P}=F|\Phi^{+}\rangle\langle\Phi^{+}|+(1-F)|\Phi^{-}\rangle\langle\Phi^{-}|.\label{logicphase}
\end{eqnarray}
 As shown in Fig. 1, they choose two copies of  mixed states of the form of Eq. (\ref{logicphase}). Therefore, the whole system can be described as follows.
 With the probability of $F^{2}$, it is in the state $|\Phi^{+}\rangle_{A_{1}B_{1}}|\Phi^{+}\rangle_{A_{2}B_{2}}$. With the equal probability of $F(1-F)$, they are in the states $|\Phi^{+}\rangle_{A_{1}B_{1}}|\Phi^{-}\rangle_{A_{2}B_{2}}$ and $|\Phi^{-}\rangle_{A_{1}B_{1}}|\Phi^{+}\rangle_{A_{2}B_{2}}$. With the probability of $(1-F)^{2}$,  they are in the state $|\Phi^{-}\rangle_{A_{1}B_{1}}|\Phi^{-}\rangle_{A_{2}B_{2}}$. We first discuss the state $|\Phi^{+}\rangle_{A_{1}B_{1}}|\Phi^{+}\rangle_{A_{2}B_{2}}$.  State  $|\Phi^{+}\rangle_{A_{1}B_{1}}|\Phi^{+}\rangle_{A_{2}B_{2}}$ can be described as
  \begin{eqnarray}
 &&|\Phi^{+}\rangle_{A_{1}B_{1}}|\Phi^{+}\rangle_{A_{2}B_{2}}\nonumber\\
 &=&\frac{1}{\sqrt{2}}(|\phi^{+}\rangle_{A_{1}}|\phi^{+}\rangle_{B_{1}}+|\phi^{-}\rangle_{A_{1}}|\phi^{-}\rangle_{B_{1}})\nonumber\\
&\otimes&\frac{1}{\sqrt{2}}(|\phi^{+}\rangle_{A_{2}}|\phi^{+}\rangle_{B_{2}}+|\phi^{-}\rangle_{A_{2}}|\phi^{-}\rangle_{B_{2}})\nonumber\\
  &=&\frac{1}{2}(|\phi^{+}\rangle_{A_{1}}|\phi^{+}\rangle_{B_{1}}|\phi^{+}\rangle_{A_{2}}|\phi^{+}\rangle_{B_{2}}\nonumber\\
  &+&|\phi^{+}\rangle_{A_{1}}|\phi^{+}\rangle_{B_{1}}|\phi^{-}\rangle_{A_{2}}|\phi^{-}\rangle_{B_{2}}\nonumber\\
  &+&|\phi^{-}\rangle_{A_{1}}|\phi^{-}\rangle_{B_{1}}|\phi^{+}\rangle_{A_{2}}|\phi^{+}\rangle_{B_{2}}\nonumber\\
  &+&|\phi^{-}\rangle_{A_{1}}|\phi^{-}\rangle_{B_{1}}|\phi^{-}\rangle_{A_{2}}|\phi^{-}\rangle_{B_{2}}).\label{state4}
 \end{eqnarray}\\

 \begin{figure}[!h]
\begin{center}
\includegraphics[width=6cm,angle=0]{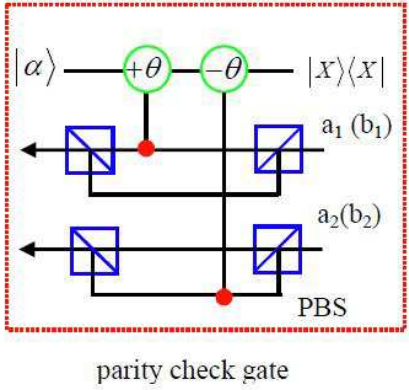}
\caption{A schematic drawing of parity check gate (PCG), which has been widely used in quantum information processing. The similar construction of PCG is also shown in Refs.\cite{concentration3,concentration4,cnot1}. It can distinguish the states $|H\rangle|H\rangle$ and $|V\rangle|V\rangle$ from $|H\rangle|V\rangle$, $|V\rangle|H\rangle$ near deterministically.}
\end{center}
\end{figure}

Similar to the above description, they pick up the case that all the coherent states make phase shifts $\theta$. The state in Eq. (\ref{state4}) will collapse to
  \begin{eqnarray}
  &\rightarrow&\frac{1}{\sqrt{2}}(|\Phi^{+}\rangle_{A_{1}B_{1}}+|\Psi^{+}\rangle_{A_{1}B_{1}})(|\Phi^{+}\rangle_{A_{2}B_{2}}+|\Psi^{+}\rangle_{A_{2}B_{2}})\nonumber\\
  &+&(|\Phi^{+}\rangle_{A_{1}B_{1}}-|\Psi^{+}\rangle_{A_{1}B_{1}})(|\Phi^{+}\rangle_{A_{2}B_{2}}-|\Psi^{+}\rangle_{A_{2}B_{2}})\nonumber\\
  &=&\frac{1}{\sqrt{2}}(|\Phi^{+}\rangle_{A_{1}B_{1}}|\Phi^{+}\rangle_{A_{2}B_{2}}+|\Psi^{+}\rangle_{A_{1}B_{1}}|\Psi^{+}\rangle_{A_{2}B_{2}}).\label{evolve7}
 \end{eqnarray}
 Certainly, states $|\Phi^{+}\rangle_{A_{1}B_{1}}|\Psi^{+}\rangle_{A_{2}B_{2}}$ and $|\Psi^{+}\rangle_{A_{1}B_{1}}|\Phi^{+}\rangle_{A_{2}B_{2}}$ cannot lead all the coherent states pick up the phase shift $\theta$, which can be eliminated automatically. The last state By selecting the case that all the coherent state pick up the phase shift $\theta$, state $|\Psi^{+}\rangle_{A_{1}B_{1}}|\Psi^{+}\rangle_{A_{2}B_{2}}$ will become
 \begin{eqnarray}
 &\rightarrow&\frac{1}{\sqrt{2}}(|\Phi^{-}\rangle_{A_{1}B_{1}}+|\Psi^{-}\rangle_{A_{1}B_{1}})(|\Phi^{-}\rangle_{A_{2}B_{2}}+|\Psi^{-}\rangle_{A_{2}B_{2}})\nonumber\\
  &+&(|\Phi^{-}\rangle_{A_{1}B_{1}}-|\Psi^{-}\rangle_{A_{1}B_{1}})(|\Phi^{-}\rangle_{A_{2}B_{2}}-|\Psi^{-}\rangle_{A_{2}B_{2}})\nonumber\\
  &=&\frac{1}{\sqrt{2}}(|\Phi^{-}\rangle_{A_{1}B_{1}}|\Phi^{-}\rangle_{A_{2}B_{2}}+|\Psi^{-}\rangle_{A_{1}B_{1}}|\Psi^{-}\rangle_{A_{2}B_{2}}).\label{evolve8}
  \end{eqnarray}
  From Eqs. (\ref{evolve7}) and (\ref{evolve8}), Alice and Bob measure the photons in spatial modes $a_{3}$, $b_{3}$, $a_{4}$ and $b_{4}$ in the basis $\{|\pm\rangle\}$.  If the measurement result is are all the same, say  $|+\rangle_{a_{3}}|+\rangle_{a_{4}}|+\rangle_{b_{3}}|+\rangle_{b_{4}}$ or
 $|-\rangle_{a_{3}}|-\rangle_{a_{4}}|-\rangle_{b_{3}}|-\rangle_{b_{4}}$, they will obtain a new mixed state $\rho''_{A_{1}B_{1}}$
 \begin{eqnarray}
\rho''_{A_{1}B_{1}}=F'|\Phi^{+}\rangle_{A_{1}B_{1}}\langle\Phi^{+}|+(1-F')|\Phi^{-}\rangle_{A_{1}B_{1}}\langle\Phi^{-}|,\label{new2}
\end{eqnarray}
  Here $F'=\frac{F^{2}}{F^{2}+(1-F)^{2}}$, if $F>\frac{1}{2}$, $F'> F$. On the other hand, if the measurement result in spatial modes  $a_{3}a_{4}$  are opposite, i. e. $|+\rangle_{a_{3}}|-\rangle_{a_{4}}$, or $|-\rangle_{a_{3}}|+\rangle_{a_{4}}$, they should perform a phase flip operation on one of the photons in $a_{1}$ or $a_{2}$ modes. If the measurement result in spatial modes $b_{3}b_{4}$ are opposite, i. e. $|+\rangle_{b_{3}}|-\rangle_{b_{4}}$, or $|-\rangle_{b_{3}}|+\rangle_{b_{4}}$, they should also perform a phase flip operation on one of the photons in $b_{1}$ or $b_{2}$ modes. In this way, they can obtain the same mixed state in Eq. (\ref{new2}). In this way, they complete the distillation.

   \section{Correction of physical bit-flip error}
   If the logic-qubit entanglement $|\Phi^{+}\rangle_{A_{1}B_{1}}$ suffer from a physical bit-flip error in physical qubit $a_{1}$ with the probability of $1-F$.
   The whole state will become a mixed state as
   \begin{eqnarray}
\rho_{PB}=F|\Phi^{+}\rangle_{A_{1}B_{1}}\langle\Phi^{+}|+(1-F)|\Upsilon^{+}\rangle_{A_{1}B_{1}}\langle\Upsilon^{+}|.\label{physicalbit}
\end{eqnarray}
Here the subscript $PB$ means the physical bit. Interestingly, such error occurs on one of the logic qubit, which is locally. Therefore, it can be corrected
completely. In this distillation, they do not require two pairs of mixed states. As shown in Fig. 2, if the error occurs on the logic qubit $A_{1}$, they let the photon $a_{1}$ and $a_{2}$ in the state
of Eq. (\ref{physicalbit}) pass through the PBSs, respectively. Interestingly, $|\phi\rangle_{a_{1}a_{2}}$ will make the coherent state $|\alpha\rangle$ pick up the phase shift $\theta$, while $|\phi\rangle_{a_{1}a_{2}}$ will make the coherent state $|\alpha\rangle$ pick us no phase shift. In this way, by check the phase shift of the state, they can easily find the bit-flip error deterministically. On the other hand, if the physical bit-flip error occurs in the logic qubit $B_{1}$, it can also be checked with the same principle. Once they find the error qubit, they only need to perform a bit-flip operation on one of the physical qubit to correct it.

\section{Distillation of arbitrary C-GHZ state }
This protocol can also be used to distill the logic-qubit entanglement with each logic qubit being arbitrary physical GHZ state.
For example, suppose Alice and Bob share the logic-qubit entanglement
\begin{eqnarray}
|\Phi^{+}_{m}\rangle&=&\frac{1}{\sqrt{2}}(|GHZ^{+}_{m}\rangle|GHZ^{+}_{m}\rangle\nonumber\\
&+&|GHZ^{-}_{m}\rangle|GHZ^{-}_{m}\rangle).
\end{eqnarray}

 \begin{figure}[!h]
\begin{center}
\includegraphics[width=8cm,angle=0]{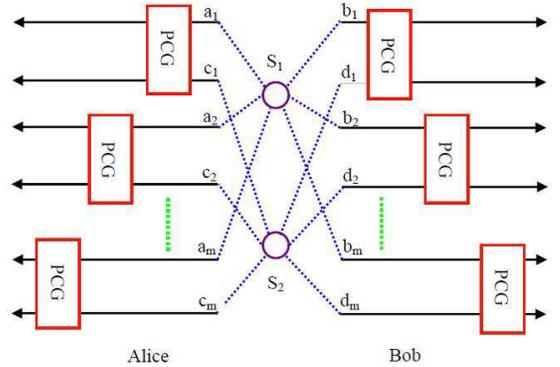}
\caption{A schematic drawing of distillation of arbitrary logic-qubt entanglement. In each logic qubit, it is an $m$-photon GHZ state.}
\end{center}
\end{figure}
The noise environment will make state  $|\Phi^{+}_{m}\rangle$ become a mixed state as
\begin{eqnarray}
\rho^{m}=F|\Phi^{+}_{m}\rangle\langle\Phi_{m}^{+}|+(1-F)|\Psi^{+}_{m}\rangle\langle\Psi_{m}^{+}|.\label{aghz}
\end{eqnarray}
Here
\begin{eqnarray}
|\Psi^{+}_{m}\rangle&=&\frac{1}{\sqrt{2}}(|GHZ^{+}_{m}\rangle|GHZ^{-}_{m}\rangle\nonumber\\
&+&|GHZ^{-}_{m}\rangle|GHZ^{+}_{m}\rangle).
\end{eqnarray}
As shown in Fig. 3, the PCG is the parity check gate in Fig. 2. Alice and Bob choose two copies of mixed state to perform distillation.
The first mixed state is in the spatial modes $a_{1}$, $a_{2}$, $\cdots$,  $a_{m}$, $b_{1}$, $b_{2}$, $\cdots$, $b_{m}$. The second mixed state is in the
spatial modes $c_{1}$, $c_{2}$, $\cdots$,  $c_{m}$, $d_{1}$, $d_{2}$, $\cdots$, $d_{m}$. The two mixed states can be described as: with the probability of
$F^{2}$, it is in the state $|\Phi^{+}_{m}\rangle_{AB}\otimes|\Phi^{+}_{m}\rangle_{CD}$. With the equal probability of $F(1-F)$, they are in the states $|\Phi^{+}_{m}\rangle_{AB}\otimes|\Psi^{+}_{m}\rangle_{CD}$ and $|\Psi^{+}_{m}\rangle_{AB}\otimes|\Phi^{+}_{m}\rangle_{CD}$. With the probability of $(1-F)^{2}$, it is in the state $|\Psi^{+}_{m}\rangle_{AB}\otimes|\Psi^{+}_{m}\rangle_{CD}$. Before distillation, they first perform the Hadamard operations on all the photons. The principle of distillation is similar to the previous approach. They select the case that all the coherent states in PCGs pick up the phase shift $\theta$. State $|\Phi^{+}_{m}\rangle_{AB}\otimes|\Phi^{+}_{m}\rangle_{CD}$ will collapse to
\begin{eqnarray}
&\rightarrow&\frac{1}{2^{m}}[(|H\rangle_{a_{1}}|H\rangle_{a_{2}}\cdots|H\rangle_{a_{m}}|H\rangle_{c_{1}}|H\rangle_{c_{2}}\cdots|H\rangle_{c_{m}}\nonumber\\
&+&|H\rangle_{a_{1}}|V\rangle_{a_{2}}\cdots|V\rangle_{a_{m}}|H\rangle_{c_{1}}|V\rangle_{c_{2}}\cdots|V\rangle_{c_{m}}\nonumber\\
&+&\cdots\nonumber\\
&+&|V\rangle_{a_{1}}|V\rangle_{a_{2}}\cdots|H\rangle_{a_{m}}|V\rangle_{b_{1}}|V\rangle_{b_{2}}\cdots|H\rangle_{b_{m}})\nonumber\\
&\otimes&(|H\rangle_{b_{1}}|H\rangle_{b_{2}}\cdots|H\rangle_{b_{m}}|H\rangle_{d_{1}}|H\rangle_{d_{2}}\cdots|H\rangle_{d_{m}}\nonumber\\
&+&|H\rangle_{b_{1}}|V\rangle_{b_{2}}\cdots|V\rangle_{b_{m}}|H\rangle_{d_{1}}|V\rangle_{d_{2}}\cdots|V\rangle_{d_{m}}\nonumber\\
&+&\cdots\nonumber\\
&+&|V\rangle_{b_{1}}|V\rangle_{b_{2}}\cdots|H\rangle_{b_{m}}|V\rangle_{d_{1}}|V\rangle_{d_{2}}\cdots|H\rangle_{d_{m}})\nonumber\\
&+&(|H\rangle_{a_{1}}|H\rangle_{a_{2}}\cdots|V\rangle_{a_{m}}|H\rangle_{c_{1}}|H\rangle_{c_{2}}\cdots|V\rangle_{c_{m}}\nonumber\\
&+&\cdots\nonumber\\
&+&|V\rangle_{a_{1}}|V\rangle_{a_{2}}\cdots|V\rangle_{a_{m}}|V\rangle_{c_{1}}|V\rangle_{c_{2}}\cdots|V\rangle_{c_{m}})\nonumber\\
&\otimes&(|H\rangle_{b_{1}}|H\rangle_{b_{2}}\cdots|V\rangle_{b_{m}}|H\rangle_{d_{1}}|H\rangle_{d_{2}}\cdots|V\rangle_{d_{m}}\nonumber\\
&+&\cdots\nonumber\\
&+&|V\rangle_{b_{1}}|V\rangle_{b_{2}}\cdots|V\rangle_{b_{m}}|V\rangle_{d_{1}}|V\rangle_{d_{2}}\cdots|V\rangle_{d_{m}})].\label{evolve9}
\end{eqnarray}
The last state $|\Psi^{+}_{m}\rangle_{AB}\otimes|\Psi^{+}_{m}\rangle_{CD}$ will collapse to
\begin{eqnarray}
&\rightarrow&\frac{1}{2^{m}}[(|H\rangle_{a_{1}}|H\rangle_{a_{2}}\cdots|H\rangle_{a_{m}}|H\rangle_{c_{1}}|H\rangle_{c_{2}}\cdots|H\rangle_{c_{m}}\nonumber\\
&+&|H\rangle_{a_{1}}|V\rangle_{a_{2}}\cdots|V\rangle_{a_{m}}|H\rangle_{c_{1}}|V\rangle_{c_{2}}\cdots|V\rangle_{c_{m}}\nonumber\\
&+&\cdots\nonumber\\
&+&|V\rangle_{a_{1}}|V\rangle_{a_{2}}\cdots|H\rangle_{a_{m}}|V\rangle_{b_{1}}|V\rangle_{b_{2}}\cdots|H\rangle_{b_{m}})\nonumber\\
&\otimes&(|H\rangle_{b_{1}}|H\rangle_{b_{2}}\cdots|V\rangle_{b_{m}}|H\rangle_{d_{1}}|H\rangle_{d_{2}}\cdots|V\rangle_{d_{m}}\nonumber\\
&+&\cdots\nonumber\\
&+&|V\rangle_{b_{1}}|V\rangle_{b_{2}}\cdots|V\rangle_{b_{m}}|V\rangle_{d_{1}}|V\rangle_{d_{2}}\cdots|V\rangle_{d_{m}})\nonumber\\
&+&(|H\rangle_{a_{1}}|H\rangle_{a_{2}}\cdots|V\rangle_{a_{m}}|H\rangle_{c_{1}}|H\rangle_{c_{2}}\cdots|V\rangle_{c_{m}}\nonumber\\
&+&\cdots\nonumber\\
&+&|V\rangle_{a_{1}}|V\rangle_{a_{2}}\cdots|V\rangle_{a_{m}}|V\rangle_{c_{1}}|V\rangle_{c_{2}}\cdots|V\rangle_{c_{m}})\nonumber\\
&\otimes&(|H\rangle_{b_{1}}|H\rangle_{b_{2}}\cdots|H\rangle_{b_{m}}|H\rangle_{d_{1}}|H\rangle_{d_{2}}\cdots|H\rangle_{d_{m}}\nonumber\\
&+&|H\rangle_{b_{1}}|V\rangle_{b_{2}}\cdots|V\rangle_{b_{m}}|H\rangle_{d_{1}}|V\rangle_{d_{2}}\cdots|V\rangle_{d_{m}}\nonumber\\
&+&\cdots\nonumber\\
&+&|V\rangle_{b_{1}}|V\rangle_{b_{2}}\cdots|H\rangle_{b_{m}}|V\rangle_{d_{1}}|V\rangle_{d_{2}}\cdots|H\rangle_{d_{m}})].\label{evolve10}
\end{eqnarray}
 States $|\Phi^{+}_{m}\rangle_{AB}\otimes|\Psi^{+}_{m}\rangle_{CD}$ and $|\Psi^{+}_{m}\rangle_{AB}\otimes|\Phi^{+}_{m}\rangle_{CD}$ cannot lead all the coherent states pick up the same phase shift $\theta$. Finally, by measuring the photons in $c_{1}$, $c_{2}$, $\cdots$, $c_{m}$, and $d_{1}$, $d_{2}$, $\cdots$, $d_{m}$ modes in the basis $|\pm\rangle$, and performing the Hadamard operations on the photons in $a_{1}$, $a_{2}$, $\cdots$, $a_{m}$, and $b_{1}$, $b_{2}$, $\cdots$, $b_{m}$ modes, they will obtain a mixed state
\begin{eqnarray}
\rho_{AB}'^{m}=F'|\Phi^{+}_{m}\rangle_{AB}\langle\Phi_{m}^{+}|+(1-F')|\Psi^{+}_{m}\rangle_{AB}\langle\Psi_{m}^{+}|,\label{aghz1}
\end{eqnarray}
if the measurement results are all the same, i. e., $|+\rangle_{c_{1}}|+\rangle_{c_{2}}\cdots|+\rangle_{c_{m}}|+\rangle_{d_{1}}|+\rangle_{d_{2}}\cdots|+\rangle_{d_{m}}$, $|-\rangle_{c_{1}}|-\rangle_{c_{2}}\cdots|-\rangle_{c_{m}}|-\rangle_{d_{1}}|-\rangle_{d_{2}}\cdots|-\rangle_{d_{m}}$ or  Here $F'=\frac{F^{2}}{F^{2}+(1-F)^{2}}$. If the measurement results are the other case, for example, it is $|+\rangle_{c_{1}}|+\rangle_{c_{2}}\cdots|-\rangle_{c_{m}}$ in Alice's location. They should perform a phase-flip operation on the state in $a_{1}$, $a_{2}$, $\cdots$, $a_{m}$ modes after performing the Hadamard operations. On the other hand, if the entanglement suffers from the logic phase-flip error, it will make state  $|\Phi^{+}_{m}\rangle$ become $|\Phi^{-}_{m}\rangle$ with the probability of $(1-F)$. The mixed state can also be distilled with the same approach. Briefly speaking, they first choose two copies of the mixed states. They let the two mixed states pass through the PCGs, subsequently. By selecting the case that all the coherent states pick up the same phase shift $\theta$, with the probability of $\frac{F^{2}}{2}$, they will obtain
\begin{eqnarray}
&\rightarrow&\frac{1}{\sqrt{2}}(|\Phi_{m}^{+}\rangle_{AB}|\Phi_{m}^{+}\rangle_{CD}+|\Psi_{m}^{+}\rangle_{AB}|\Psi_{m}^{+}\rangle_{CD}).\nonumber\\\label{evolve11}
\end{eqnarray}
With the probability of $\frac{(1-F)^{2}}{2}$, they will obtain
\begin{eqnarray}
&\rightarrow&\frac{1}{\sqrt{2}}(|\Phi_{m}^{-}\rangle_{AB}|\Phi_{m}^{-}\rangle_{CD}+|\Psi_{m}^{-}\rangle_{AB}|\Psi_{m}^{-}\rangle_{CD}).\nonumber\\\label{evolve12}
\end{eqnarray}
Finally, with the same step, they measure the photons in  $c_{1}$, $c_{2}$, $\cdots$, $c_{m}$, and $d_{1}$, $d_{2}$, $\cdots$, $d_{m}$ modes in basis $|+\rangle$. They can ultimately obtain a new mixed state with the fidelity of $F'$. Finally, let us briefly discuss the correction of physical bit-flip error.
If the physical bit-flip error occurs on the logic qubit $A$ and make the state $|\Phi_{m}^{+}\rangle_{AB}$ becomes
\begin{eqnarray}
|\Upsilon_{m}^{+}\rangle_{AB}=\frac{1}{\sqrt{2}}(|\Omega_{m}^{+}\rangle_{A}|GHZ_{m}^{+}\rangle_{B}+|\Omega_{m}^{-}\rangle_{A}|\phi^{-}\rangle_{B}).\label{aghz2}
\end{eqnarray}
Here
 \begin{eqnarray}
|\Omega_{m}^{+}\rangle_{A}&=&\frac{1}{\sqrt{2}}(|V\rangle_{a_{1}}|H\rangle_{a_{2}}|H\rangle_{a_{3}}\cdots|H\rangle_{a_{m}}\nonumber\\
&+&|H\rangle_{a_{1}}|V\rangle_{a_{2}}|V\rangle_{a_{2}}\cdots|V\rangle_{a_{m}}).
\end{eqnarray}
The physical bit-flip error can be easily checked using the PCG in Fig. 2. They let the photons in $a_{1}$ and $a_{2}$ pass through the PCG. If the coherent state shows no phase shift, they perform a bit-flip operation on the photon in $a_{1}$ or $a_{2}$. If the physical bit-flip error occurs on the other photon, it can also be checked with the similar approach. They are only required to let the error photon and its neighbor photon to pass through the PCG and measure the phase of the coherent state.

\section{Discussion and conclusion}
So far, we have completely described this protocol. We first explain this protocol with the case that each logic qubit is the two-photon polarization Bell state. We mainly described the  distillation of three kinds of errors. The first is logic bit-flip error. The second is the logic phase-flip error and the third is the physical bit-flip error. We show that the physical phase-flip error equals to the logic bit-flip error. We also extend this protocol to distill the arbitrary logic-qubit entanglement for each logic qubit being the arbitrary $m$-photon GHZ state.

 During the whole protocol, they all select the case that all the coherent states pick up the phase shift $\theta$. Actually, such selection condition is to pick up the even parity states $|H\rangle|H\rangle$ or  $|V\rangle|V\rangle$, which makes the whole protocol extremely low.  Other selection cases can also perform the successful distillation. As shown in Fig. 1,  we denote the even parity check in $a_{1}a_{2}$ as $E_{a_{1}a_{2}}$, the odd parity check as  in $a_{1}a_{2}$ as $O_{a_{1}a_{2}}$. The section condition for all the coherent states picking up phase shift $\theta$ can  be written as $E_{a_{1}a_{2}}E_{b_{1}b_{2}}E_{a_{3}a_{4}}E_{b_{3}b_{4}}$. The other success distillation can be written as $O_{a_{1}a_{2}}O_{b_{1}b_{2}}O_{a_{3}a_{4}}O_{b_{3}b_{4}}$,  $E_{a_{1}a_{2}}E_{b_{1}b_{2}}O_{a_{3}a_{4}}O_{b_{3}b_{4}}$ and $O_{a_{1}a_{2}}O_{b_{1}b_{2}}E_{a_{3}a_{4}}E_{b_{3}b_{4}}$, respectively. We take the case $O_{a_{1}a_{2}}O_{b_{1}b_{2}}O_{a_{3}a_{4}}O_{b_{3}b_{4}}$ for example. In this case, state $|\Phi^{+}\rangle_{A_{1}B_{1}}|\Phi^{+}\rangle_{A_{2}B_{2}}$ will collapse to
  \begin{eqnarray}
&&\frac{1}{4}[(|H\rangle_{a_{1}}|H\rangle_{a_{2}}|V\rangle_{a_{3}}|V\rangle_{a_{4}}+|V\rangle_{a_{1}}|V\rangle_{a_{2}}|H\rangle_{a_{3}}|H\rangle_{a_{4}})\nonumber\\
&&\otimes(|H\rangle_{b_{1}}|H\rangle_{b_{2}}|V\rangle_{b_{3}}|V\rangle_{b_{4}}+|V\rangle_{b_{1}}|V\rangle_{b_{2}}|H\rangle_{b_{3}}|H\rangle_{b_{4}}).\nonumber\\
&+&(|H\rangle_{a_{1}}|V\rangle_{a_{2}}|V\rangle_{a_{3}}|H\rangle_{a_{4}}+|V\rangle_{a_{1}}|H\rangle_{a_{2}}H\rangle_{a_{3}}|V\rangle_{a_{4}})\nonumber\\
&&\otimes(|H\rangle_{b_{1}}|V\rangle_{b_{2}}|V\rangle_{b_{3}}|H\rangle_{b_{4}}
+|V\rangle_{b_{1}}|H\rangle_{b_{2}}H\rangle_{b_{3}}|V\rangle_{b_{4}})],\label{result3}\nonumber\\
\end{eqnarray}
and state  $|\Psi^{+}\rangle_{A_{1}B_{1}}|\Psi^{+}\rangle_{A_{2}B_{2}}$  will collapse to
\begin{eqnarray}
&&\frac{1}{4}[(|H\rangle_{a_{1}}|H\rangle_{a_{2}}|V\rangle_{a_{3}}|V\rangle_{a_{4}}+|V\rangle_{a_{1}}|V\rangle_{a_{2}}|H\rangle_{a_{3}}|H\rangle_{a_{4}})\nonumber\\
&\otimes&(|H\rangle_{b_{1}}|V\rangle_{b_{2}}|V\rangle_{b_{3}}|H\rangle_{b_{4}}+|V\rangle_{b_{1}}|H\rangle_{b_{2}}|H\rangle_{b_{3}}|V\rangle_{b_{4}})\nonumber\\
&+&(|H\rangle_{a_{1}}|V\rangle_{a_{2}}|V\rangle_{a_{3}}|H\rangle_{a_{4}}+|V\rangle_{a_{1}}|H\rangle_{a_{2}}|H\rangle_{a_{3}}|V\rangle_{a_{4}})\nonumber\\
&\otimes&(|H\rangle_{b_{1}}|H\rangle_{b_{2}}|V\rangle_{b_{3}}|V\rangle_{b_{4}}+|V\rangle_{b_{1}}|V\rangle_{b_{2}}|H\rangle_{b_{3}}|H\rangle_{b_{4}})].\label{result4}\nonumber\\
\end{eqnarray}
Compared with Eq. (\ref{result1}) and  Eq. (\ref{result3}), Eq. (\ref{result2}) and  Eq. (\ref{result4}), they can perform two bit-flip operations on the photons $a_{3}a_{4}$ and $b_{3}b_{4}$ to convert the state in Eq. (\ref{result3}) to Eq. (\ref{result1}) and Eq. (\ref{result4}) to  Eq. (\ref{result2}), respectively.

During the protocol, we exploit the cross-Kerr nonlinearity to complete the task. We require that they can distinguish and measurement the phase shift $\theta$ in a single-photon level. The early research shows that the phase shift in a single-photon level is extremely low and it is impossible to be observed in experiment \cite{addkerr}. Recently, the important progresses in both theory and experiment showed that it is possible to obtain the large phase shift in an observable value \cite{kerr7,kerr8,kerr9}. For instance, Hoi \emph{et al.} showed that the average  phase shift of cross-Kerr was demonstrated up to 20 degrees per photon with both coherent microwave fields at the single-photon level \cite{kerr7}. Feizpour \emph{et al.} reported the first direct measurement of the cross-phase shift due to single photons \cite{kerr9}. This experiment  opens a door to future studies of nonlinear optics in the quantum regime. It also provides the  potential applications in quantum information processing.

In conclusion, we have described an approach of distilling the logic-qubit entanglement. We mainly explained the method of distilling three kinds of errors.
The first is the logic bit-flip error. The second is the logic phase-flip error and the third is the physical bit-flip error. The physical phase-flip error equals to logic bit-flip error. During the protocol, we exploit the feasible cross-Kerr nonlinearity to complete the distillation. We also extend this protocol to the case of the logic-qubit entanglement with each logic qubit being the arbitrary GHZ state.
 As the logic-qubit entanglement is an alternative resource for long-distance quantum communication,  this protocol may have its potential application in future.

\section*{ACKNOWLEDGEMENTS}
This work is supported by the National Natural Science Foundation of
China under Grant No. Grant Nos. 11474168 and 61401222, the Natural Science Foundation of
Jiangsu under Grant No. BK20151502, the Qing Lan Project in Jiangsu Province, and a Project Funded by the Priority
Academic Program Development of Jiangsu Higher Education
Institutions.

\end{document}